\documentclass{article}                    
\newcommand{\hpaa}{\hat{P}^{a}_{\alpha}}
\newcommand{\hpaae}{\hat{P}^{a}_{\alpha_1}}
\newcommand{\hpaat}{\hat{P}^{a}_{\alpha_2}}
\newcommand{\hpaas}{\hat{P}^{a}_{\alpha_3}}
\newcommand{\hpbb}{\hat{P}^{b}_\beta}
\newcommand{\hpbbl}{\hat{P}^{b}_{\beta_l}}

\newcommand{\hpaai}{\hat{P}^{a}_{\alpha_i}}
\newcommand{\hpaaj}{\hat{P}^{a}_{\alpha_j}}
\newcommand{\hpaak}{\hat{P}^{a}_{\alpha_k}}
\newcommand{\Paa}{{P}^{a}_{\alpha}}
\newcommand{\Paae}{{P}^{a}_{\alpha_1}}
\newcommand{\Paat}{{P}^{a}_{\alpha_2}}
\newcommand{\Paas}{{P}^{a}_{\alpha_3}}
\newcommand{\Paai}{{P}^{a}_{\alpha_i}}
\newcommand{\Paaj}{{P}^{a}_{\alpha_j}}
\newcommand{\Paak}{{P}^{a}_{\alpha_k}}

\newcommand{\paai}{{p}^{a}_{\alpha_i}}
\newcommand{\paaj}{{p}^{a}_{\alpha_j}}

\newcommand{\Pbb}{{P}^{b}_\beta}
\newcommand{\hpnae}{{\hat{P}^{a\bot}_{\alpha_1}}}
\newcommand{\hpnat}{{\hat{P}^{a\bot}_{\alpha_2}}}
\newcommand{\hpnas}{{\hat{P}^{a\bot}_{\alpha_3}}}

\newcommand{\ebb}{e^b_{\beta}}
\newcommand{\eaa}{e^a_{\alpha}}
\newcommand{\ebbl}{e^b_{\beta_l}}

\newcommand{\eaam}{e^a_{\alpha_m}}
\newcommand{\eaai}{e^a_{\alpha_i}}
\newcommand{\eaaj}{e^a_{\alpha_j}}
\newcommand{\eaak}{e^a_{\alpha_k}}
\newcommand{\bra}[1]{\left\langle{#1}\right\vert}
\newcommand{\ket}[1]{\left\vert{#1}\right\rangle}

\usepackage{mathptmx}      
\usepackage{latexsym}
\usepackage{verbatim}
\usepackage{amsfonts}
\usepackage{amsmath}
\title{On hyperbolic interferences in the quantum--like representation algorithm for the case of triple--valued observables
}

\author{Peter Nyman\\International Center for Mathematical Modeling\\
in Physics and Cognitive Sciences,\\
Linnaeus University, S-35195, Sweden \\\\
              {peter.nyman@lnu.se} 
}

\begin{document}
\maketitle

\begin{abstract}
The quantum--like representation algorithm (QLRA) was  introduced by A. Khrennikov \cite{K1,K2,K3,K4,K5} to solve the ``inverse Born's rule problem'', i.e. to construct a representation of probabilistic data-- measured in any context of science-- and represent this data by a complex or more general\footnote{A Clifford algebra is introduced for this more general representation} probability amplitude which matches a generalization of Born's rule. 
The outcome from QLRA will introduce the formula of total probability with an additional term of trigonometric, hyperbolic or hyper-trigonometric interference and this is in fact a generalization of the familiar formula of interference of probabilities.

We study representation of statistical data (of any origin) by a probability amplitude in a complex algebra and a Clifford algebra (algebra of hyperbolic numbers). The statistical datas are collected from measurements of two trichotomous observables and the complexity of the problem increased eventually compared to the case of dichotomous observables.  We see that only special statistical data (satisfying a number of nonlinear constraints) have a quantum--like representation.
In this paper we will present a class of statistical data which satisfy these nonlinear constraints and have a quantum--like representation.
This quantum--like representation induces trigonometric-, hyperbolic- and hyper--trigonometric interferences representation.
\end{abstract}

\section{Introduction}

The inter-relation between classical and quantum probabilistic data was discussed in numerous papers 
(from various points of view), see\footnote{The list of references is far 
from complete, see Khrennikov's monographs \cite{K6,K7} for a detailed list of references.}, e.g., \cite{VN,GD1,GD2,GD3,BL,DV,NAN,NAN1,AL,K6}. We are interested in the representation of probabilistic 
data and of any origin\footnote{Thus it need not be produced by quantum measurements; it can be collected in e.g. psychology,
see \cite{K9}.} by a generalization of complex probability amplitudes (wave function). This problem was discussed in detail in \cite{K7}. Here we describe a class of probabilistic data which permits the quantum-like (QL)
representation for the case of  two trichotomous observable. Recently the interest to trichotomous observable increased in connection with experiments of Greger Weihs' groups to check validity of Born's rule for triple slit experiment \cite{Urbasi}. This test was proposed by R.D. Sorkin \cite{S1,S2} and this is an important test of foundation of quantum mechanics. Experimental studies \cite{Urbasi} are characterized by essential increasing of complexity compared to the well known two slit experiment. We met the same increasing of complexity in our general theoretical study.      

Many papers have been published concerning inter-relation between psychology and quantum-like representation, e.g., \cite{basieva2011dynamics,busemeyer2006quantum,conte2009mental,conte2007some,model2011quantum}.
Busemeyer et al. give examples of experiments from cognitive psychology where the formula of total probabilities fails. They suggest the use of a quantum-like model \cite{busemeyer2009empirical}.
In the experiment students look at pictures that depict people with narrow width and
thick lips or wide width and thin lips. The students tasks are to categorize the faces belonging to either a `good' or `bad' guy group and also judge if the person have a withdraw or attack action.
This can be written as probabilities of categorizing that a face belong to a `good' or `bad' guy group respectively. Similarly we have the probabilities representing withdraw or attack action. The experiment also give the conditional probabilities, e.g., attack action conditional to `good' guy. They show that the law of total probability is violated for some of this collected probability data. To match this data, which is a violation of the law of total probability, they introduce quantum-like models. In \cite{K9} A. Khrennikov presented the hypothesis that biological systems might use complex probabilistic amplitudes (``mental wave functions'') in processing of statistical data.
If this hypothesis is correct, then these amplitudes can be reconstructed on the basis of 
collected experimental data. In psychology this approach got the name ``constructive wave function approach''.

A general QL-representation algorithm (QLRA) was presented in \cite{K7}. This algorithm is based 
on the {\it  formula of total probability with interference term} -- a disturbance of the standard formula of total 
probability. Starting with experimental probabilistic data, QLRA produces a complex probability amplitude such that 
probability can be reconstructed by using Born's rule. 

Although the formal scheme of QLRA works for multi-valued observables of an arbitrary dimension, the description of the class of 
probabilistic data which can be transfered into QL-amplitudes (the domain of application of QLRA)  
depends very much on the dimension. In \cite{N1} the simplest case of data generated by dichotomous observables was studied. Examples of probabilistic data that generate QL-amplitudes are studied in \cite{IrinaPeter_article}. The complexity of the problem increases incredibly compared to the two dimensional case, but here we show a class of probabilistic data that generate QL-amplitudes.

 It appears naturally to represent quantum physics by hyperbolic numbers (also known as perplex, unipodal, duplex  or split-complex algebra) and  have application to physics \cite{antonuccio1998hyperbolic}. Hucks shows that a four-component Dirac spinor is equivalent to
a two-component hyperbolic complex spinor and that Lorentz group is homomorphic to the hyperbolic unitary group \cite{hucks1993hyperbolic}. Ulrych analysis the symmetry of the hyperbolic Hilbert space and representation of Poincar{\'e} mass operator in the hyperbolics algera \cite{ulrych2005poincaré}.
 
The scheme of presentation is the following. We start with observables given by quantum mechanics (QM) and derive 
constraints on phases which are necessary and sufficient for the QL-representation. Then we use 
these constraints to produce complex amplitudes and more general (so called hyperbolic) amplitudes from data (of any origin); some examples, including numerical, are 
given. 
In this paper we stress the role of hyperbolic amplitudes, i.e., amplitudes valued in a special Clifford algebra, so called hyperbolic algebra, see e.g. \cite{Kisil}. 

\section{Trichotomous incompatible quantum observables}

\subsection{Probabilities}
\label{HH7}

Let $\hat{a}$ and $\hat{b}$ be two self-adjoint operators in three dimensional complex Hilbert space representing two trichotomous incompatible observables $a$ and $b$. They take values $a=\alpha_i,\;i=1,2,3$ and $b=\beta_l, l=1,2,3$ --
spectra of operators. We assume that the operators have non-degenerated spectra, i.e., $\alpha_i \not=\alpha_j, \beta_i \not=\beta_j, i \not=j.$
Consider corresponding eigenvectors: $$\hat{a} e^{a}_{\alpha_i}={\alpha_i} e^{a}_{\alpha_i}, \quad \hat{b} e^{b}_{\beta_l}={\beta_l} e^{b}_{\beta_l}.$$
Denote by $\hpaai=\ket{\eaai}\bra{\eaai}$ and $\hpbbl=\vert{\ebbl}\rangle\langle{\ebbl}\vert$ one dimensional projection operators and by $\Paa$ and $\Pbb$ the observables represented by there projections. Consider also projections \begin{align}
\hpnae=\hpaat+\hpaas,\quad \hpnat=\hpaae+\hpaas,\quad \hpnas=\hpaae+\hpaat.
\end{align}
Here the observable $\Paai=1$ if the result of the $a$-measurement is $a=\alpha_i$ and $\Paai=0$ if  $a\not=\alpha_i.$
The  observables $\Pbb$ are defined in the same way.
We have the following relation between  events corresponding to measurements
\begin{align}
[\Paae=0]&=[\Paat=1]\vee[\Paas=1], \quad [\Paat=0]=[\Paae=1]\vee[\Paas=1],\\\nonumber
[\Paas=0]&=[\Paae=1]\vee[\Paat=1].
\end{align}
Here the probabilities given by the QM-formalism are 
\begin{align}\label{prob1} p^b_{\beta}&\equiv P_\psi (b=\beta)= ||\hpbb \psi||^2=|\langle{\psi} \vert{\ebb}\rangle|^2,
\\\nonumber p^a_{\alpha}&\equiv P_\psi (a=\alpha)=||\hpaa \psi||^2=|\langle{\psi}\vert {\eaa}\rangle|^2,
\end{align}
where $\psi$ is a wave function.
We also have the conditional (transition) probabilities given by the QM-formalism as 
\begin{equation}
p^{b|a}_{\beta\alpha} \equiv  P_\psi(b=\beta|\Paa=1)=||\hpbb \hpaa \psi||^2/||\hpaa\psi||^2=|\langle\eaa|\ebb\rangle|^2.
\end{equation} 
 
We remark that non-degeneration of the spectra implies that they do not depend on $\psi.$  
Moreover, the matrix of transition probabilities is {\it doubly stochastic}.
We will require the following conditions (compare this with classical probability theory) for these conditional probabilities;
\begin{equation}\label{sumcon}
\sum_{i=1}^3 p^{b|a}_{\beta_i\alpha_j}=1,
\end{equation} 
for all $j=1,2,3$.
Here the $\psi$-dependent conditional probabilities are 
\begin{align}
p_{\beta_l\alpha_{kj}}^{b|a} &\equiv P_\psi(b=\beta_l|\Paak = 1\vee \Paaj = 1)= P_\psi(b=\beta_l|\Paai=0)\\\nonumber &=\frac{\left\vert\right\vert\hpbbl(\hpaaj+\hpaak)\psi\left\vert\right\vert^2}{\left\vert\right\vert
(\hpaaj+\hpaak)\psi\left\vert\right\vert^2},
\end{align} 
where $\;i,j,k\in \{1,2,3\},\; j,k  \neq i.$
We have 
\begin{align}
\frac{\left\vert\right\vert\hpbb(\hpaaj+\hpaak)\psi\left\vert\right\vert^2}{\left\vert\right\vert(\hpaaj+\hpaak)\psi\left\vert\right\vert^2}&=
\frac{\left\vert\right\vert\vert{\ebb}\rangle\langle\ebb\vert{\eaaj}\rangle\langle{\eaaj}\vert\psi\rangle+\vert{\ebb}\rangle\langle\ebb\vert{\eaak}\rangle\langle{\eaak}\vert\psi\rangle\left\vert\right\vert^2}{\left\vert\right\vert\vert{\eaaj}\rangle\langle{\eaaj}\vert\psi\rangle+\vert{\eaak}\rangle\langle{\eaak}\vert\psi\rangle\left\vert\right\vert^2}
\\\nonumber
&=\frac{\left\vert\right\vert\vert{\ebb}\rangle\left\vert\right\vert^2\vert\langle\ebb\vert{\eaaj}\rangle\langle{\eaaj}\vert\psi\rangle+\langle\ebb\vert{\eaak}\rangle\langle{\eaak}\vert\psi\rangle\vert^2}{\vert\langle{\eaaj}\vert\psi\rangle\vert^2+\vert\langle{\eaak}\vert\psi\rangle\vert^2}
\\\nonumber
&=\frac{\vert\langle\ebb\vert{\eaaj}\rangle\langle{\eaaj}\vert\psi\rangle+\langle\ebb\vert{\eaak}\rangle\langle{\eaak}\vert\psi\rangle\vert^2}{\vert\langle{\eaaj}\vert\psi\rangle\vert^2+\vert\langle{\eaak}\vert\psi\rangle\vert^2}.
\end{align}
Note that $p_{\beta_l\alpha_{kj}}^{b|a}=p_{\beta_l\alpha_{jk}}^{b|a}$.

\section{Clifford algebra (hyperbolic algebra)}\label{Clifford algebra }
As mention before we will consider the complex Hilbert space but also the hyperbolic Hilbert space.
Therefore let us define a Clifford algebra called hyperbolic algebra (see \cite{K7}) with the purpose to define the hyperbolic Hilbert space.
The formalism for this hyperbolic algebra is similar to conventional complex numbers. This algebra contains expressions as unit circle, Euler's formula and conjugate.
Thus, let an element $z$ belong to the hyperbolic algebra ${\bf G}$ if and only if it have following form;
\[z=x+j y,\quad x,y \in \mathbb{R},\]
where $j^2=1$,\; $z_1+z_2=x_1+x_2+j(y_1+y_2)$ and  
$z_1 z_2=x_1 x_2+y_1+y_2+j(y_1 x_2+y_2 x_1)$.
This algebra is a commutative two-dimensional algebra with two orthonormal basis $e_0=1$ and $e_1=1$. The hyperbolic conjugate is defined as $\bar{z}=x-j y$ where obviously $\bar{z}\in {\bf G}$.
Moreover, the square of absolute value is defined by
\[|z|^2=z \bar{z}=x^2-y^2\] and 
therefore will $|z|^2\in{\bf G}$ , in fact $|z|^2\in\mathbb{R}$.
But $|z|$ will not be well defined for $z$ such that $|z|^2\leq0$.
Therefore set \[\mathbf{G}_+=\{z\in\mathbf G:|z|^2\geq0\}.\]
and 
\[\mathbf{G}^*_+=\{z\in\mathbf G:|z|^2>0\}.\]
Moreover define the argument $\arg{z}$ for $z\in \mathbf{G^*_+}$ as
\[\arg{z}=\operatorname{arctanh}{\frac{y}{x}}=\frac{1}{2}\ln\frac{x+y}{x-y}.\]
Notice that $x\neq0,\;x-y\neq0$ and $\frac{x+y}{x-y}>0$,\;since $z \in \mathbf{G}^*_+$.
We also define a hyperbolic exponential function 
\[e^{j \theta}=\cosh \theta+j \sin\theta, \quad \theta\in\mathbb{R}.\]
Since $\cosh \theta \geq0$, elements $z\in G_+^*$ with $x<0$ can not be represented by $|z|e^{j \theta}$.
Therefore, in order to represent all elements $z\in G_+^*$ put \[z=\epsilon|z|e^{j\theta},\]  where $\epsilon=x/|x|$ and $\arg z= \theta$.
Moreover, this is a multiplicative semigroup.
Let $z_1,z_2\in \mathbf{G}^*_+$, so $|z_1|^2,|z_2|^2 >0$
then we see that $z_1\cdot z_2\in \mathbf{G}^*_+$ by $$|z_1  z_2|^2=|\epsilon_1|z_1|e^{j\theta_1}\epsilon_2|z_2|e^{j\theta_2}|^2=|z_1|^2|z_2|^2>0.$$
But, when we add two elements $z_1,z_2\in \mathbf{G}^*_+$ it follows that the existence of elements so that
$z_1+z_2 \not\in \mathbf{G}^*_+.$
Let us analyze for which of the elements $z_1,z_2\in \mathbf{G}^*_+$, $z_1+z_2 \not\in \mathbf{G}^*_+$, i.e. $|z_1+z_2|^2 <0$.
It follows that \begin{align}\label{add}|z_1+z_2|^2&=\big|\epsilon_1|z_1|e^{j\theta_1}+\epsilon_2|z_2|e^{j\theta_2}\big|^2\\\nonumber
&=|z_1|^2+|z_2|^2+2\epsilon_1\epsilon_2|z_1||z_2|\cosh(\theta_1-\theta_2)\end{align}
From \eqref{add} and  $\cosh(\theta_1-\theta_2)>0$ it follows that $|z_1+z_2|^2>0$ if $\epsilon_1\epsilon_2=1$.
Here we consider elements $z_1,z_2\in \mathbf{G}^*_+$ such that $|z_1|^2+|z_2|^2+2\epsilon_1\epsilon_2|z_1||z_2|\cosh(\theta_1-\theta_2)>0$.
Let $\epsilon_1\epsilon_2=-1$ then $|z_1+z_2|^2>0$ if and only if $$\operatorname{arccosh}\left(\frac{|z_1|^2+|z_2|^2}{2|z_1||z_2|}\right)>|\theta_1-\theta_2|.$$

\subsection{Hyperbolic Hilbert space}

A hyperbolic Hilbert space $H$ is a $\mathbf{G}$-linear inner
product space. Let $x,y,z\in H$ and $a,b\in \mathbf{G}$, then
consider the inner product as a map from $H\times H$ to
$\mathbf{G}$ having the following properties:

 \medskip
(1) Conjugate symmetry: $\left\langle x,y\right\rangle$ is the
conjugate to $\left\langle y,x\right\rangle$
\[\left\langle x,y\right\rangle=\overline{\left\langle y,x \right\rangle}\]

(2) Linearity with respect to  the first argument:
\[\left\langle a x +b z,y\right\rangle=a\left\langle x,y\right\rangle+b\left\langle z,y\right\rangle\]

(3) Non-degenerate:
  \[\left\langle x,y\right\rangle=0\]
for all $y\in H$ if and only if $x=0$ \\
In general, the norm $\left\|\psi \right\|=\sqrt{\left\langle \psi,
\psi\right\rangle}$ is not well defined. But we only need  the
square of the norm $\left\|\psi \right\|^2=\left\langle \psi, \psi
\right\rangle$.

\subsection{Probability amplitudes}

Set 
$
\psi_{\beta} = \langle\psi |\ebb\rangle.
$ and consider the complex Hilbert space and the hyperbolic Hilbert space.
Then by Born's rule 
\begin{equation}
\label{RRTT}
p_\beta^b = \vert \psi_\beta\vert^2.
\end{equation}
We have 
\begin{equation}
\label{RRTT1}
\psi = \sum_\beta \psi_\beta \ebb.
\end{equation}
Thus these amplitudes give a possibility to reconstruct the state.
We remark that $\psi= \sum_\alpha \hpaa \psi$, hence
\begin{equation}
\label{RRTT2}
\psi_{\beta}= \sum_\alpha \langle \hpaa \psi |\ebb\rangle.
\end{equation}
Each amplitude $\psi_{\beta}$ can be represented as the sum of subamplitudes
\begin{equation}
\label{RRTT3}
\psi_{\beta}= \sum_\alpha \psi_{\beta\alpha}
\end{equation}
 given by 
\begin{equation}\label{RRTT4}\psi_{\beta\alpha}=\langle\hpaa\psi|\ebb\rangle=\langle\psi|\eaa\rangle\langle\eaa|\ebb\rangle.
\end{equation} 
Hence, one can reconstruct the state $\psi$ on the basis of amplitudes $\psi_{\beta\alpha}.$
We remark that
$$|\psi_{\beta_l\alpha_i}|^2=|\langle\psi|\eaai\rangle\langle\eaai|\ebbl\rangle|^2=p^a_{\alpha_i} p_{\beta_l\alpha_i}^{b|a}.
$$
In this notations
\begin{equation}\label{pblakj} p_{\beta_l\alpha_{kj}}^{b|a}=|\psi_{\beta_l\alpha_k}+\psi_{\beta_l\alpha_j}|^2/(p^a_{\alpha_j}+p^a_{\alpha_k}).
\end{equation}
Here $|\psi_{\beta_l\alpha_i}|=\sqrt{p^{a}_{\alpha_i}p^{b|a}_{\beta_l,\alpha_i}}$ 
and therefore  \begin{equation}
\label{tio1}\psi_{\beta_l\alpha_i}=\sqrt{p^{a}_{\alpha_i}p^{b|a}_{\beta_l,\alpha_i}}\lambda_{ \varphi_{\beta_l\alpha_i}},
\end{equation} where $|\lambda_{ \varphi_{\beta_l\alpha_i}}|=1.$
Moreover, put 
\begin{equation}
\label{tio}
\langle \psi|\eaam\rangle
=\sqrt{p_{\alpha_m}^a}\lambda_{ \xi_{\alpha_m}}, \; \langle \eaam|\ebbl\rangle =\sqrt{p_{\beta_l \alpha_m}^{ab}}\lambda_{ \theta_{\beta_l\alpha_m}},
\end{equation}
where $|\lambda_{ \xi_{\alpha_m}}|=1$ and $|\lambda_{ \theta_{\beta_l\alpha_m}}|=1.$
Hence, it follows from \eqref{RRTT4} and \eqref{tio1} that 
\begin{equation}\label{tioH}
\lambda_{ \varphi_{\beta_l\alpha_i}}=\lambda_{ \xi_{\alpha_m}}\lambda_{ \theta_{\beta_l\alpha_m}}.
\end{equation} 
We have a system of equations for phases $\psi_{\beta_l\alpha_i}$  for $i,j,k,l \in\{1,2,3\}$,
\begin{align}
\label{atta1} 
|\psi_{\beta_l\alpha_i}+\psi_{\beta_l\alpha_j}|^2&=\vert \langle\psi|\eaai\rangle\langle\eaai|\ebbl\rangle+\langle\psi|\eaaj\rangle\langle\eaaj|\ebbl\rangle|^2
\\\nonumber
&=|\langle\psi|\eaai\rangle\langle\eaai|\ebbl\rangle|^2+|\langle\psi|\eaaj\rangle\langle\eaaj|\ebbl\rangle|^2
\\\nonumber
&+ \langle\psi|\eaai\rangle \langle\eaai|\ebbl\rangle \langle\ebbl|\eaaj\rangle \langle\eaaj|\psi\rangle
\\\nonumber
&+\langle\eaai|\psi\rangle \langle\ebbl|\eaai\rangle \langle\psi|\eaaj\rangle\langle\eaaj|\ebbl\rangle
\\\nonumber
&=p^{a}_{\alpha_i}p^{b|a}_{\beta_l,\alpha_i}+p^{a}_{\alpha_j}p^{b|a}_{\beta_l,\alpha_j}
\\\nonumber &+2\lambda_{l,ij}\sqrt{p^{a}_{\alpha_i}p^{b|a}_{\beta_l,\alpha_i}p^{a}_{\alpha_j}p^{b|a}_{\beta_l,\alpha_j}}
\end{align}
where we put \begin{equation}\label{lambdalij}\lambda_{l,ij}\equiv\frac{1}{2}\left(\lambda_{ \varphi_{\beta_l\alpha_i}}\overline{\lambda_{ \varphi_{\beta_l\alpha_j}}}+\lambda_{ \varphi_{\beta_l\alpha_j}}\overline{\lambda_{ \varphi_{\beta_l\alpha_i}}}\right).
\end{equation}
Thus, the coefficients of interference $\lambda_{l,ij}$ can be written by \eqref{pblakj} and \eqref{atta1} as \begin{equation}\label{sju}
\lambda_{l,ij}\equiv\frac{(\paai+\paaj)p_{\beta_l\alpha_{ij}}^{b|a}-(\paai p_{\beta_l\alpha_{i}}^{b|a}+\paaj p_{\beta_l\alpha_{j}}^{b|a})}{2\sqrt{\paai p_{\beta_l\alpha_{i}}^{b|a}\paaj p_{\beta_l\alpha_{j}}^{b|a}}}.
\end{equation}
\subsection{Interference Classification}
Note that the coefficients of interference in \eqref{sju} will take values in $\mathbb{R}$. 
We divide this into following two cases of interference depending on the absolute value of $\lambda_{l,ij}$.
\\
\begin{enumerate}
	\item Let $|\lambda_{l,ij}|\leq 1$ and in this case put $\lambda_{ \varphi_{\beta_l\alpha_i}}=e^{i  \varphi_{\beta_l\alpha_i}}$, then it straight forward from \eqref{tioH} that $e^{i  \varphi_{\beta_l\alpha_i}}=e^{i(   \xi_{\alpha_i}+\theta_{\beta_l\alpha_i})}$.
Thus, by \eqref{lambdalij} we see that $$\lambda_{l,ij}=\cos(\varphi_{\beta_l\alpha_i}-\varphi_{\beta_l\alpha_j}).$$
We refer to this interference as trigonometric interference.
	\item Let  $|\lambda_{l,ij}|> 1$ and put $\lambda_{ \varphi_{\beta_l\alpha_i}}=\epsilon_{l,i}e^{j  \varphi_{\beta_l\alpha_i}}$ 
	where $j^2=1$ and $\epsilon_{l,i}={\lambda_{ \varphi_{\beta_l\alpha_i}}}/{|\lambda_{ \varphi_{\beta_l\alpha_i}}|}$. Here the symbol $j$ is a generator of the Clifford algebra $\mathbf{G}^*_+=\{z\in\mathbf G:|z|^2>0\}$, (let us call it Hyperbolic algebra). An element $z\in \mathbf{G}^*_+$ have the form $z=x+j y$ and the ``hyperbolic conjugate''\footnote{Please note that this is not the usual complex conjugate! This is the analogous conjugate associated to the Clifford algebra.
We are aware of the fact that this might be confusing, but would still like to call this conjugate because of its similarity to the complex conjugate.} $z=x-j y$. It is apparent that $\bar {z}\in{\bf G}^*_+$.
 	We introduce the hyperbolic exponential function
\begin{equation}\label{HEF}
e^{j \theta}=\cosh \theta+j \sinh\theta, \quad \theta \in \mathbb{R}.
\end{equation}
We also use the identities
\begin{equation}\label{EF2}
\cosh\theta=\frac{e^{j \theta}+e^{-j \theta}}{2}\quad \text{and} \quad \sinh\theta=\frac{e^{j \theta}-e^{-j \theta}}{2j}.
\end{equation}
By \eqref{lambdalij} it follows that 
$$\lambda_{l,ij}=\epsilon_{l,ij}\cosh(\varphi_{\beta_l\alpha_i}-\varphi_{\beta_l\alpha_j}),$$ where $\epsilon_{l,ij}=\epsilon_{l,i} \epsilon_{l,j}$.
We will call this hyperbolic interference.
\end{enumerate}
\subsection{Formula of total probability with interference term}

By using the definition of the amplitude $\psi_{\beta_l\alpha_i}=\langle\psi|\eaa\rangle\langle\eaa|\ebb\rangle$ we obtain
\begin{align}
\label{atta} 
p_{\beta_l}^b&= |\psi_{\beta_l\alpha_i}+\psi_{\beta_l\alpha_j}+\psi_{\beta_l\alpha_k}|^2
\\\nonumber
&=\vert \langle\psi|\eaai\rangle\langle\eaai|\ebbl\rangle+\langle\psi|\eaaj\rangle\langle\eaaj|\ebbl\rangle
+\langle\psi|\eaak\rangle\langle\eaak|\ebbl\rangle|^2
\\\nonumber
&=|\langle\psi|\eaai\rangle\langle\eaai|\ebbl\rangle|^2+|\langle\psi|\eaaj\rangle\langle\eaaj|\ebbl\rangle|^2
+|\langle\psi|\eaak\rangle\langle\eaak|\ebbl\rangle|^2+
\\\nonumber
&+ \langle\psi|\eaai\rangle \langle\eaai|\ebbl\rangle \langle\ebbl|\eaaj\rangle \langle\eaaj|\psi\rangle
+ \langle\psi|\eaai\rangle \langle\eaai|\ebbl\rangle \langle\ebbl|\eaak\rangle \langle\eaak|\psi\rangle
\\\nonumber
&+ \langle\psi|\eaaj\rangle \langle\eaaj|\ebbl\rangle \langle\ebbl|\eaak\rangle \langle\eaak|\psi\rangle
+\langle\eaai|\psi\rangle \langle\ebbl|\eaai\rangle \langle\psi|\eaaj\rangle\langle\eaaj|\ebbl\rangle
\\\nonumber
&+\langle\eaai|\psi\rangle \langle\ebbl|\eaai\rangle \langle\psi|\eaak\rangle\langle\eaak|\ebbl\rangle
+\langle\eaaj|\psi\rangle \langle\ebbl|\eaaj\rangle \langle\psi|\eaak\rangle\langle\eaak|\ebbl\rangle .
\end{align}
Finally, we obtain  
\begin{align}\label{elva} p_{\beta_l}^b
&=p_{\alpha_i}^a p_{\beta_l \alpha_i}^{b|a}+p_{\alpha_j}^a p_{\beta_l \alpha_j}^{b|a}+p_{\alpha_k}^a p_{\beta_l \alpha_k}^{b|a}
+2\lambda_{l,ij}\sqrt{p_{\alpha_i}p_{\alpha_j}p_{\beta_l \alpha_i}^{b|a}p_{\beta_l \alpha_j}^{b|a}}
\\\nonumber&+2\lambda_{l,ik}\sqrt{p_{\alpha_i}p_{\alpha_k}p_{\beta_l \alpha_i}^{b|a}p_{\beta_l \alpha_k}^{b|a}}
+2\lambda_{l,jk}\sqrt{p_{\alpha_j}p_{\alpha_k}p_{\beta_l \alpha_j}^{b|a}p_{\beta_l \alpha_k}^{b|a}}.
\end{align}
Here, if $|\lambda_{l,ij}|\leq1$ for all $l,i,j=1,2,3,\; i\neq j$ then we call this the case of trigonometric interference and the case where $|\lambda_{l,ij}|>1$ for all $l,i,j=1,2,3,\; i\neq j$ is called the case of hyperbolic interference. All other cases are combinations of these two cases of interference and is called the hyper-trigonometric interference (i.e. for some $l,i,j=1,2,3,\; i\neq j$, $|\lambda_{l,ij}|\leq1$ and for the rest $|\lambda_{l,ij}|>1$ ).  
Equation \eqref{elva} is nothing else than the formula of total probability with the interference terms. It can be considered \cite{K10}  as a perturbation 
of the classical formula of total probability
\begin{equation} 
\label{elva1} 
p_{\beta_l}^b =p_{\alpha_i}^a p_{\beta_l \alpha_i}^{b|a}+p_{\alpha_j}^a p_{\beta_l \alpha_j}^{b|a}+p_{\alpha_k}^a p_{\beta_l \alpha_k}^{b|a}.
\end{equation}
If all coefficients of interferences $\lambda_{l.ij}=0$, then \eqref{elva} coincides with \eqref{elva1}.
\subsection{Unitarity of transition operator}
\label{Un}

We now remark that bases consisting of $\hat{a}$- and $\hat{b}$-eigenvectors are orthogonal; hence the operator of transition from
one basis to another is unitarity. We can always select the $b$-basis in the canonical way 
\begin{equation}
\label{L1}
e^b_{\beta_1}=\left(
\begin{array}{ll}
 1\\
 0\\
 0
 \end{array}
 \right ), 
 \; e^b_{\beta_2}=\left(
\begin{array}{ll}
 0\\
 1\\
 0
 \end{array}
 \right ),\; 
e^b_{\beta_3}=\left(
\begin{array}{ll}
 0\\
 0\\
 1
 \end{array}
 \right ).
\end{equation}
In this system of coordinates the $a$-basis has the form
\begin{align}
\label{L2}
e^a_{\alpha_1}&=\left(
\begin{array}{ll}
 \sqrt{p_{\beta_1 \alpha_1}}\lambda_{ \varphi_{\beta_1\alpha_1}}\\
 \sqrt{p_{\beta_2 \alpha_1}}\lambda_{ \varphi_{\beta_2 \alpha_1}}\\
\sqrt{p_{\beta_3 \alpha_1}}\lambda_{ \varphi_{\beta_3 \alpha_1}}
 \end{array}
 \right ), 
\quad 
e^a_{\alpha_2}=\left(
\begin{array}{ll}
\sqrt{p_{\beta_1 \alpha_2}}\lambda_{ \varphi_{\beta_1 \alpha_2}}\\
\sqrt{p_{\beta_2 \alpha_2}}\lambda_{ \varphi_{\beta_2 \alpha_2}} \\
\sqrt{p_{\beta_3 \alpha_2}}\lambda_{ \varphi_{\beta_3 \alpha_2}}
 \end{array}
 \right ), 
\\\nonumber
e^a_{\alpha_3}&=\left(
\begin{array}{ll}
\sqrt{p_{\beta_1 \alpha_3}}\lambda_{ \varphi_{\beta_1 \alpha_3}}\\
\sqrt{p_{\beta_2 \alpha_3}}\lambda_{ \varphi_{\beta_3 \alpha_3}}\\
\sqrt{p_{\beta_3 \alpha_3}}\lambda_{ \varphi_{\beta_3 \alpha_3}}
 \end{array}
 \right ). 
\end{align}
The matrix
$$
U =\left(
\begin{array}{lll}
 \sqrt{p_{\beta_1 \alpha_1}} \lambda_{ \varphi_{\beta_1 \alpha_1}} \; \sqrt{p_{\beta_1 \alpha_2}} \lambda_{ \varphi_{\beta_1 \alpha_2}}\; 
\sqrt{p_{\beta_1 \alpha_3}} \lambda_{ \varphi_{\beta_1 \alpha_3}} \\
\sqrt{p_{\beta_2 \alpha_1}} \lambda_{ \varphi_{\beta_2 \alpha_1}}\; \sqrt{p_{\beta_2 \alpha_2}} \lambda_{ \varphi_{\beta_2 \alpha_2}}\; 
\sqrt{p_{\beta_2 \alpha_3}} \lambda_{ \varphi_{\beta_2 \alpha_3}} \\
 \sqrt{p_{\beta_3 \alpha_1}} \lambda_{ \varphi_{\beta_3 \alpha_1}}\; \sqrt{p_{\beta_3 \alpha_2}} \lambda_{ \varphi_{\beta_3 \alpha_2}}\; 
\sqrt{p_{\beta_3 \alpha_3}} \lambda_{ \varphi_{\beta_3 \alpha_3}}
 \end{array}
 \right)
$$ 
is unitary. Hence, we have the system of equations 
\begin{equation}
\label{UIZ}
 \sum_m \sqrt{p_{\beta_m \alpha_i}p_{\beta_m \alpha_k}} \lambda_{\varphi_{\beta_m \alpha_i}} \overline{\lambda_{\varphi_{\beta_m \alpha_k}}}= 0
 \end{equation}
 and 
\begin{equation}
\label{UIZ1}
 \sum_m p_{\beta_m \alpha_i} |\lambda_{\varphi_{\beta_m \alpha_i}}|^2= 1,
 \end{equation} 
 from the condition that $U$ is unitary.\\
 The second equations \eqref{UIZ1} will always hold  by \eqref{sumcon} and $|\lambda_{\varphi_{\beta_m \alpha_i}}|=1$ ,see \eqref{tio1}.\\
The first of this equations \eqref{UIZ} can be rewritten by \eqref{tioH} as 
 \begin{equation}
\label{UIZ 2}
\lambda_{\xi_{\alpha_i}}\overline{\lambda_{\xi_{\alpha_k}}}\sum_m \sqrt{p_{\beta_m \alpha_i}p_{\beta_m \alpha_k}} \lambda_{\theta_{\beta_m \alpha_i}} \overline{\lambda_{\theta_{\beta_m \alpha_k}}}= 0,
 \end{equation}
 or
  \begin{equation}\label{tn}
 \sum_m \sqrt{p_{\beta_m \alpha_i}p_{\beta_m \alpha_k}} \lambda_{\theta_{\beta_m \alpha_i}} \overline{\lambda_{\theta_{\beta_m \alpha_k}}}= 0,
 \end{equation}
 where $\lambda_{\xi_{\alpha_i}},\lambda_{\xi_{\alpha_k}}\neq0.$\\
 Thus  \eqref{tn} imply unitarity of $U$.
\section{Selection of orthonormal bases for the unitarity of transition operator}
We now show by example that there exist unitary $U$ satisfying QLRA. Let us choose to work with the case of hyperbolic interference. The same calculations with similar basis can be done with trigonometric interference. When repeating these calculations for trigonometric interference put $\lambda_{ \varphi_{\beta_l\alpha_i}}=e^{i  \theta_{\beta_l\alpha_i}}$.
Choose orthonormal $a$-bases by setting $\lambda_{ \varphi_{\beta_l\alpha_i}}=\epsilon_{li}e^{j  \theta_{\beta_l\alpha_i}}, p_{\alpha_k\beta_i}=a^2_{k i}/(1+a^2_{2 i}+a^2_{3 i}),\;\epsilon_{1i}=1,\; a_{1 i}=1,\; \theta_{\beta_1 \alpha_i}=u,\;\theta_{\beta_2 \alpha_i}=s ,\;\theta_{\beta_3 \alpha_i}=t,\;i=1,2,3$ in \eqref{L2}; 
$$
\label{EXL2}
e^a_{\alpha_i}=\frac{1}{\sqrt{1+a_{2i}^2+a_{3i}^2}}\left(
\begin{array}{c}
  e^{j u}\\
 \epsilon_{2i}a_{2i} e^{j s}\\
 \epsilon_{3i}a_{3i} e^{j t}
 \end{array}
 \right ). 
$$
Since the bases are orthonormal we have that 
$$\left\langle e_{\alpha_i}|e_{\alpha_k}\right\rangle=
\begin{cases} \frac{1+a_{2 i} a_{2 k} \epsilon _{2 i} \epsilon _{2 k}+a_{3 i} a_{3 k} \epsilon _{3 i} \epsilon _{3 k}}{\sqrt{(a_{2i}^2+a_{3i}^2+1)(a_{2k}^2+a_{3k}^2+1)}}=0,& \text{if $k\neq i$,}
\\
1 &\text{if $i=k$.}
\end{cases}
$$
where $\;i,k=1,2,3.$
The case $k\neq i$ give us a system of equations
$1+a_{2 i} a_{2 k} \epsilon _{2 i} \epsilon _{2 k}+a_{3 i} a_{3 k} \epsilon _{3 i} \epsilon _{3 k}=0,\;i,k=1,2,3,$
with solution 
\begin{align} \label{sol}a_{31}&=\frac{-a_{23}^2 \epsilon _{23}^2-a_{32} a_{33} \epsilon _{32} \epsilon _{33}-1}{\epsilon _{31} \left(a_{23}^2 a_{32} \epsilon _{32} \epsilon
   _{23}^2+a_{32} a_{33}^2 \epsilon _{32} \epsilon _{33}^2+a_{33} \epsilon _{33}\right)}
,\\ \nonumber a_{21}&= -\frac{a_{23} \epsilon _{23} \left(a_{32} \epsilon _{32}-a_{33}
   \epsilon _{33}\right)}{\epsilon _{21} \left(a_{23}^2 a_{32} \epsilon _{32} \epsilon _{23}^2+a_{32} a_{33}^2 \epsilon _{32} \epsilon _{33}^2+a_{33} \epsilon
   _{33}\right)}
   ,\\ \nonumber a_{22}&=\frac{-a_{32} a_{33} \epsilon _{32} \epsilon _{33}-1}{a_{23} \epsilon _{22} \epsilon _{23}}.
   \end{align}
   Then let \begin{equation}\label{psi1}
   \psi=\frac{1}{\sqrt{v_1^2+v_2^2+v_3^2}}(v_1 e^{\gamma _1 j}e_{\beta_1}+v_2 e^{j\gamma _2}e_{\beta_2}+v_3 e^{j \gamma _3}e_{\beta_3}).\end{equation} Note that $| \psi|^2 = 1$.
  Next we calculate the probabilities  $p_{\alpha_i}=| \langle{\eaai}\vert{\psi}\rangle|^2$ for $i=1,2,3$, in consideration of solutions \eqref{sol}. In order to reduce the size of the expressions we introduce the following rewriting. 
\begin{align}\label{cpro}
    d_1&=\left(1+a_{23}^2+a_{32} a_{33} \epsilon _{32} \epsilon _{33}\right),
  &d_2&=\left(a_{32} \left(a_{23}^2+a_{33}^2\right) \epsilon _{32}+a_{33} \epsilon _{33}\right),
  \\\nonumber d_3&=a_{23}^2 \left(a_{32} \epsilon _{32}-a_{33} \epsilon _{33}\right){}^2,
  &d_4&=a_{23}\epsilon _{23} \left(a_{32} \epsilon _{32}-a_{33} \epsilon _{33}\right),
  \\\nonumber d_5&=\left(1+a_{32} a_{33} \epsilon _{32} \epsilon _{33}\right),
  &d_6&=a_{32} v_3 \epsilon _{32},
  \\\nonumber d_7&=a_{23} v_2 \epsilon _{23},
  &d_8&=a_{33} v_3 \epsilon _{33},
   \\\nonumber d_9&=a_{23}^2 a_{32}^2 v_3^2,
   &d_{10}&=\left(a_{23} \epsilon _{23}+a_{23} a_{32} a_{33} \epsilon _{23} \epsilon _{32} \epsilon _{33}\right){}^2
   \\\nonumber d_{11}&=1+a_{23}^2+a_{33}^2
   &d_{12}&=a_{23}^2 \left(1+a_{32}^2\right)+d_5^2
    \\\nonumber d_{13}&=v_1^2+v_2^2+v_3^2
  \\\nonumber \gamma _{\text{s12}}&= \gamma _1-\gamma _2+s-u,
  &\gamma _{\text{t13}}&= \gamma _1-\gamma _3+t-u
 \end{align}
We then get the following probabilities,
%
%
\begin{align*}p_{\alpha_1}&=\big(d_2^2 v_1^2-2 d_2 \left(\cosh \left(\gamma _{\text{s12}}\right) d_4 v_2+\cosh \left(\gamma _{\text{t13}}\right) d_1 v_3\right) v_1+d_3 v_2^2\\\nonumber &+d_1 v_3
   \left(2 \cosh \left(\gamma _{\text{s12}}-\gamma _{\text{t13}}\right) d_4 v_2+d_1 v_3\right)\big)\big({d_{11} d_{12} d_{13}}\big)^{-1},
\\
p_{\alpha_2}&=\big(\left(v_1^2+2 \cosh \left(\gamma _{\text{t13}}\right) d_6 \epsilon _{23}^2 v_1+a_{32}^2 v_3^2\right) a_{23}^2\\\nonumber &-2 d_5 \left(\cosh \left(\gamma
   _{\text{s12}}-\gamma _{\text{t13}}\right) d_6+\cosh \left(\gamma _{\text{s12}}\right) v_1\right) v_2 \epsilon _{23} a_{23}+d_5^2 v_2^2\big)\big(d_{12} d_{13}\big)^{-1},
\\
p_{\alpha_3}&=\big(v_1^2+2 \cosh \left(\gamma _{\text{t13}}\right) d_8 v_1+a_{23}^2 v_2^2+a_{33}^2 v_3^2\\&+2 d_7 \left(\cosh \left(\gamma _{\text{s12}}-\gamma
   _{\text{t13}}\right) d_8+\cosh \left(\gamma _{\text{s12}}\right) v_1\right)\big)\big(d_{11} d_{13}\big)^{-1}
   \end{align*}
The calculations of the conditional probabilities $p_{\beta_l\alpha_{kj}}^{b|a}$ are found in the appendix.
%
%
Moreover 
\begin{equation*}
p_{\beta_1\alpha_{1}}^{b|a}=\frac{d_2^2}{d_{11} d_{12}}
,\quad p_{\beta_1\alpha_{2}}^{b|a}=\frac{1}{a_{32}^2+\frac{d_5^2}{a_{23}^2}+1}
,\quad p_{\beta_1\alpha_{3}}^{b|a}=\frac{1}{d_{11}}
\end{equation*}
%
%
\begin{equation*}
p_{\beta_2\alpha_{1}}^{b|a}=\frac{d_3}{d_{11} d_{12}}
,\quad p_{\beta_2\alpha_{2}}^{b|a}=\frac{d_5^2}{d_{12}}
,\quad p_{\beta_3\alpha_{3}}^{b|a}=\frac{a_{23}^2}{d_{11}}
\end{equation*}
%
%
\begin{equation*}
p_{\beta_3\alpha_{1}}^{b|a}=\frac{d_1^2}{d_{11} d_{12}}
,\quad p_{\beta_3\alpha_{2}}^{b|a}=\frac{a_{32}^2}{a_{32}^2+\frac{d_5^2}{a_{23}^2}+1}
,\quad p_{\beta_3\alpha_{3}}^{b|a}=\frac{a_{33}^2}{d_{11}}
\end{equation*}
Please note that we require that these probabilities satisfy $ 0< p_{\beta_l\alpha_{kj}}^{b|a}<1,\;0<p_{\alpha_i}<1$. Since these probabilities also satisfy that $\sum_i^3 p_{\alpha_i}=1,\;  \sum_l^3 p_{\beta_l\alpha_{kj}}^{b|a}=1$, it is sufficient to show that $\sum_i^3 p_{\alpha_i}>0$ and  $\sum_l^3 p_{\beta_l\alpha_{kj}}^{b|a}>0$ for all $k,j$.
%
In order to show that there exist at least one such orthonormal basis in \eqref{EXL2} and $\psi$ in \eqref{psi1} that meets these requirements on the probabilities, we here give an example.
Let $\gamma _1=0,\; \gamma _2=t,\; \gamma_3=t,\;v_1=-2,\;v_2=3,\;v_3=-2,   $ in the quantum state in \eqref{psi1}
\begin{equation}\label{psiN1}
   \psi=\frac{1}{\sqrt{17}}(-2 e_{\beta_1}+3 e^{j t}e_{\beta_2}-2 e^{j t}e_{\beta_3}).
   \end{equation}

Let $\epsilon_{23}a_{23}=2,\; \epsilon_{32}a_{32}=2,\;\epsilon_{33}a_{33}=3 ,\;s=t ,\;u=0.3$ in the basis in \eqref{EXL2} 
\begin{align*}
\label{EXLN2}
e^a_{\alpha_i}=\frac{1}{\sqrt{1+\left(\frac{2}{29}\right)^2+\left(\frac{11}{29}\right)^2}}\left(
\begin{array}{c}
 e^{ 0.3 j} \\
 \frac{2}{29}e^{t j} \\
 -\frac{11}{29}e^{t j}
\end{array}
\right)&, 
\;
e^a_{\alpha_2}=\frac{1}{\sqrt{1+\left(\frac{7}{2}\right)^2+(2)^2}}\left(
\begin{array}{c}
 e^{ 0.3 j} \\
 -\frac{7}{2}e^{ t j} \\
 2e^{ t j}
\end{array}
\right),
\\ 
e^a_{\alpha_3}=\frac{1}{\sqrt{1+(2)^2+(3)^2}}&\left(
\begin{array}{c}
 e^{ 0.3 j} \\
 2e^{ t j} \\
 3e^{ t j}
\end{array}
\right).
\end{align*}%
Then $$p_{\alpha_1}=0.045837,\quad p_{\alpha_2}=0.937356,\quad  p_{\alpha_3}=0.016807,$$
$$p_{\beta_1\alpha_{12}}=0.206349,\quad p_{\beta_1\alpha_{13}}=0.887593,
\quad p_{\beta_1\alpha_{23}}=0.075727,$$$$p_{\beta_2\alpha_{12}}=0.650559,
\quad p_{\beta_2\alpha_{13}}=0.111601,\quad p_{\beta_2\alpha_{23}}=0.580032,$$$$ p_{\beta_3\alpha_{12}}=0.143091,\;p_{\beta_3\alpha_{13}}=0.000805,p_{\beta_3\alpha_{23}}=0.344240.$$
\section{Appendix: }\label{app }
Here we calculate the conditional probabilities 
$$p_{\beta_l\alpha_{kj}}^{b|a}=|\psi_{\beta_l\alpha_k}+\psi_{\beta_l\alpha_j}|^2/(p^a_{\alpha_j}+p^a_{\alpha_k}).$$
We use \eqref{cpro} to reduce the size of the expressions 
%
%
{\scriptsize
\begin{align*}p_{\beta_1\alpha_{12}}^{b|a}&=-\frac{1}{d_{11}}+1-\bigg(a_{33}^2 v_2^2+a_{23}^2 v_3^2-2 \cosh \left(\gamma _{\text{s12}}-\gamma _{\text{t13}}\right) d_7 d_8\bigg)\\&\bigg[\left(v_1^2+v_3^2\right) a_{23}^2+v_2^2+v_3^2-2
   \cosh \left(\gamma _{\text{t13}}\right) d_8 v_1\\&-2 d_7 \left(\cosh \left(\gamma _{\text{s12}}-\gamma _{\text{t13}}\right) d_8+\cosh \left(\gamma
   _{\text{s12}}\right) v_1\right)+a_{33}^2 \left(v_1^2+v_2^2\right)\bigg]^{-1}
\end{align*}
 %
%
\begin{align*}p_{\beta_1\alpha_{13}}^{b|a}&=\Bigg(2\left(\left(a_{23}^2+a_{33}^2\right) a_{32}^2+2 d_5-1\right) v_1 \epsilon _{23} \left(\cosh \left(\gamma _{\text{s12}}\right) d_5 v_2 -\cosh \left(\gamma
   _{\text{t13}}\right) a_{23} d_6 \epsilon _{23}\right) a_{23}\\&-2 \cosh \left(\gamma _{\text{s12}}-\gamma _{\text{t13}}\right) d_5 d_6 v_2 \epsilon _{23} a_{23}^3+\left(d_9^2+d_5^2 v_2^2\right) a_{23}^2\\&+\left(\left(a_{23}^2+a_{33}^2\right) a_{32}^2+2 d_5-1\right){}^2 v_1^2
\Bigg)
  \\& \Bigg[d_{12} \big(\big(\left(a_{23}^2+a_{33}^2\right) a_{32}^2+2 d_5-1\big) v_1^2+a_{23}^2 \left(a_{32}^2+1\right) v_2^2+\left(a_{23}^2+d_5^2\right) v_3^2\\&+2
   a_{23} \epsilon _{23} \left(d_5 \left(\cosh \left(\gamma _{\text{s12}}-\gamma _{\text{t13}}\right) d_6+\cosh \left(\gamma _{\text{s12}}\right) v_1\right)
   v_2-\cosh \left(\gamma _{\text{t13}}\right) a_{23} d_6 v_1 \epsilon _{23}\right)\big)\Bigg]^{-1}
\end{align*}
%
%
\begin{align*}p_{\beta_1\alpha_{23}}^{b|a}&=\frac{a_{23}^2}{d_{12}}+\frac{1}{d_{11}}-\bigg(d_1^2 v_2^2-2 \cosh \left(\gamma _{\text{s12}}-\gamma _{\text{t13}}\right) d_1 d_4 v_3 v_2+d_3
   v_3^2\bigg)\\&\bigg[\left(a_{23}^4+\left(a_{32}^2+a_{33}^2+2\right) a_{23}^2+d_5^2\right) v_1^2+2 d_2 \left(\cosh \left(\gamma _{\text{s12}}\right) d_4 v_2+\cosh
   \left(\gamma _{\text{t13}}\right) d_1 v_3\right) v_1\\&+\left(\left(a_{32}^2+1\right) a_{23}^4+\left(a_{33}^2+1\right) d_5^2+2 d_{10}\right) v_2^2-2 \cosh
   \left(\gamma _{\text{s12}}-\gamma _{\text{t13}}\right) d_1 d_4 v_2 v_3\\&+v_3^2 \left(2 a_{32} \epsilon _{32} \epsilon _{33} a_{33}^3+\left(a_{23}^2+1\right)
   a_{33}^2+a_{32}^2 \left(a_{23}^4+\left(2 a_{33}^2+1\right) a_{23}^2+a_{33}^4\right)\right)\bigg]^{-1}
 \end{align*} 
  %
%
 \begin{align*}p_{\beta_2\alpha_{12}}^{b|a}&=\bigg(\left(v_1^2+2 \cosh \left(\gamma _{\text{t13}}\right) d_8 \epsilon _{23}^2 v_1+a_{33}^2 v_3^2\right) a_{23}^2\\&-2 \left(a_{33}^2+1\right) \left(\cosh
   \left(\gamma _{\text{s12}}-\gamma _{\text{t13}}\right) d_8+\cosh \left(\gamma _{\text{s12}}\right) v_1\right) v_2 \epsilon _{23}
   a_{23}+\left(a_{33}^2+1\right){}^2 v_2^2\bigg)\\&\bigg[d_{11} \left(\left(v_1^2+v_3^2\right) a_{23}^2+v_2^2+v_3^2-2 \cosh \left(\gamma _{\text{t13}}\right) d_8 v_1\right)-\\&d_{11} \left(2 d_7
   \left(\cosh \left(\gamma _{\text{s12}}-\gamma _{\text{t13}}\right) d_8+\cosh \left(\gamma _{\text{s12}}\right) v_1\right)+a_{33}^2
   \left(v_1^2+v_2^2\right)\right)\bigg]^{-1}
  \end{align*} 
  %
%
 \begin{align*}p_{\beta_2\alpha_{13}}^{b|a}&=
 \bigg(a_{23}^2 \bigg(\left(v_1^2+2 \cosh \left(\gamma _{\text{t13}}\right) d_6 v_1+a_{32}^2 v_3^2\right) d_5^2\\&+2 \left(a_{32}^2+1\right) d_7 \left(\cosh
   \left(\gamma _{\text{s12}}-\gamma _{\text{t13}}\right) d_6+\cosh \left(\gamma _{\text{s12}}\right) v_1\right) d_5+a_{23}^2 \left(a_{32}^2+1\right){}^2
   v_2^2\bigg)\bigg)\\&\bigg[d_{12} \big(\left(\left(a_{23}^2+a_{33}^2\right) a_{32}^2+2 d_5-1\right) v_1^2+a_{23}^2 \left(a_{32}^2+1\right)
   v_2^2+\left(a_{23}^2+d_5^2\right) v_3^2\\&+2 a_{23} \epsilon _{23} \big(d_5 \left(\cosh \left(\gamma _{\text{s12}}-\gamma _{\text{t13}}\right) d_6+\cosh
   \left(\gamma _{\text{s12}}\right) v_1\right) v_2-\cosh \left(\gamma _{\text{t13}}\right) a_{23} d_6 v_1 \epsilon _{23}\big)\big)\bigg]^{-1}
  \end{align*} 
   %
%
   \begin{align*}p_{\beta_2\alpha_{23}}^{b|a} &=\bigg(\left(\left(a_{32}^2+1\right) a_{23}^4+\left(a_{33}^2+1\right) d_5^2+2 d_{10}\right){}^2 v_2^2\\&-2 \cosh \left(\gamma _{\text{s12}}-\gamma
   _{\text{t13}}\right) d_1 d_4 \left(\left(a_{32}^2+1\right) a_{23}^4+\left(a_{33}^2+1\right) d_5^2+2 d_{10}\right) v_3 v_2\\&+2 d_2 d_4 v_1 \left(\cosh
   \left(\gamma _{\text{s12}}\right) \left(\left(a_{32}^2+1\right) a_{23}^4+\left(a_{33}^2+1\right) d_5^2+2 d_{10}\right) v_2-\cosh \left(\gamma
   _{\text{t13}}\right) d_1 d_4 v_3\right)\\&+d_3 \left(d_2^2 v_1^2+d_1^2 v_3^2\right)\bigg)\\&\bigg[d_{11} d_{12} \Big(\left(a_{23}^4+\left(a_{32}^2+a_{33}^2+2\right)
   a_{23}^2+d_5^2\right) v_1^2+2 d_2 \big(\cosh \left(\gamma _{\text{s12}}\right) d_4 v_2\\&+\cosh \left(\gamma _{\text{t13}}\right) d_1 v_3\big)
   v_1+\left(\left(a_{32}^2+1\right) a_{23}^4+\left(a_{33}^2+1\right) d_5^2+2 d_{10}\right) v_2^2\\&-2 \cosh \left(\gamma _{\text{s12}}-\gamma _{\text{t13}}\right)
   d_1 d_4 v_2 v_3\\&+v_3^2 \left(2 a_{32} \epsilon _{32} \epsilon _{33} a_{33}^3+\left(a_{23}^2+1\right) a_{33}^2+a_{32}^2 \left(a_{23}^4+\left(2
   a_{33}^2+1\right) a_{23}^2+a_{33}^4\right)\right)\Big)\bigg]^{-1}
  \end{align*}
   \begin{align*}p_{\beta_3\alpha_{12}}^{b|a}&=\bigg(a_{23}^2 \left(a_{23}^2 v_1^2+d_5^2 v_2^2\right) a_{32}^2+2 a_{23} \epsilon _{23} \epsilon _{32} \Big(\left(a_{23}^2+d_5^2\right) v_3 \big(\cosh
   \left(\gamma _{\text{s12}}-\gamma _{\text{t13}}\right) d_5 v_2\\&-\cosh \left(\gamma _{\text{t13}}\right) a_{23} v_1 \epsilon _{23}\big)
   -\cosh \left(\gamma
   _{\text{s12}}\right) a_{23} a_{32} d_5 d_7 v_1 \epsilon _{23} \epsilon _{32}\Big) a_{32}+\left(a_{23}^2+d_5^2\right){}^2 v_3^2\bigg)\\&\bigg[d_{12}
   \Big(\left(\left(a_{23}^2+a_{33}^2\right) a_{32}^2+2 d_5-1\right) v_1^2+a_{23}^2 \left(a_{32}^2+1\right) v_2^2+\left(a_{23}^2+d_5^2\right) v_3^2\\&+2 a_{23}
   \epsilon _{23} \left(d_5 \left(\cosh \left(\gamma _{\text{s12}}-\gamma _{\text{t13}}\right) d_6+\cosh \left(\gamma _{\text{s12}}\right) v_1\right) v_2-\cosh
   \left(\gamma _{\text{t13}}\right) a_{23} d_6 v_1 \epsilon _{23}\right)\Big)\bigg]^{-1}
  \end{align*}
   \begin{align*}p_{\beta_3\alpha_{13}}^{b|a}&=\bigg(a_{23}^2 \left(a_{23}^2 v_1^2+d_5^2 v_2^2\right) a_{32}^2+2 a_{23} \epsilon _{23} \epsilon _{32} \Big(\left(a_{23}^2+d_5^2\right) v_3 \big(\cosh
   \left(\gamma _{\text{s12}}-\gamma _{\text{t13}}\right) d_5 v_2\\&-\cosh \left(\gamma _{\text{t13}}\right) a_{23} v_1 \epsilon _{23}\big)-\cosh \left(\gamma
   _{\text{s12}}\right) a_{23} a_{32} d_5 d_7 v_1 \epsilon _{23} \epsilon _{32}\Big) a_{32}+\left(a_{23}^2+d_5^2\right){}^2 v_3^2\bigg)\\&\bigg[d_{12}
   \Big(\left(\left(a_{23}^2+a_{33}^2\right) a_{32}^2+2 d_5-1\right) v_1^2+a_{23}^2 \left(a_{32}^2+1\right) v_2^2\\&+\left(a_{23}^2+d_5^2\right) v_3^2+2 a_{23}
   \epsilon _{23} \big(d_5 \left(\cosh \left(\gamma _{\text{s12}}-\gamma _{\text{t13}}\right) d_6+\cosh \left(\gamma _{\text{s12}}\right) v_1\right) v_2
   \\&-\cosh\left(\gamma _{\text{t13}}\right) a_{23} d_6 v_1 \epsilon _{23}\big)\Big)\bigg]^{-1}
  \end{align*}
   \begin{align*}p_{\beta_3\alpha_{23}}^{b|a}&=\bigg(d_{12} \Big(\left(d_2^2 v_1^2-2 \cosh \left(\gamma _{\text{s12}}\right) d_2 d_4 v_2 v_1+d_3 v_2^2\right) d_1^2+2 \Big(\cosh \left(\gamma
   _{\text{t13}}\right) d_2 v_1
   \\&-\cosh \left(\gamma _{\text{s12}}-\gamma _{\text{t13}}\right) d_4 v_2\Big) v_3 \Big(a_{32}^2 a_{23}^4+\left(\left(2
   a_{33}^2+1\right) a_{32}^2+a_{33}^2\right) a_{23}^2+a_{33}^2\\&+a_{32} a_{33}^3 \big(a_{32} a_{33}+2 \epsilon _{32} \epsilon _{33}\big)\Big) d_1\\&+v_3^2
   \left(a_{32}^2 a_{23}^4+\left(\left(2 a_{33}^2+1\right) a_{32}^2+a_{33}^2\right) a_{23}^2+a_{33}^2+a_{32} a_{33}^3 \left(a_{32} a_{33}+2 \epsilon _{32}
   \epsilon _{33}\right)\right){}^2\Big)\bigg)
   \\&\bigg[\left(\left(a_{32}^2+1\right) a_{23}^2+d_5^2\right){}^2 d_{11} \Big(\left(d_{12} v_2^2+d_{11} \left(v_1^2+2 \cosh
   \left(\gamma _{\text{t13}}\right) d_6 \epsilon _{23}^2 v_1+a_{32}^2 v_3^2\right)\right) a_{23}^2\\&-2 d_5 d_{11} \left(\cosh \left(\gamma _{\text{s12}}-\gamma
   _{\text{t13}}\right) d_6+\cosh \left(\gamma _{\text{s12}}\right) v_1\right) v_2 \epsilon _{23} a_{23}\\&+d_5^2 d_{11} v_2^2+2 \cosh \left(\gamma
   _{\text{t13}}\right) d_8 d_{12} v_1+2 d_7 d_{12} \Big(\cosh \left(\gamma _{\text{s12}}-\gamma _{\text{t13}}\right) d_8+\cosh \left(\gamma
   _{\text{s12}}\right) v_1\Big)\\&+d_{12} \left(v_1^2+a_{33}^2 v_3^2\right)\Big)\bigg]^{-1}
  \end{align*}
}

\end{document}